\documentclass{article}

\usepackage{arxiv}

\usepackage[utf8]{inputenc}
\usepackage[T1]{fontenc}
\usepackage{microtype}

\usepackage{amsmath,amssymb,amsfonts}
\usepackage{amsthm}

\usepackage{graphicx}
\usepackage{subcaption}
\usepackage{booktabs}
\usepackage{placeins}

\usepackage{algorithm}
\usepackage{algorithmic}

\usepackage{nicefrac}
\usepackage{xcolor}

\usepackage{hyperref}
\usepackage[capitalize,noabbrev]{cleveref}

\usepackage{natbib}

\newtheorem{theorem}{Theorem}[section]

\theoremstyle{definition}

\theoremstyle{remark}
\newtheorem{remark}[theorem]{Remark}

\crefname{subsection}{subsection}{subsections}
\Crefname{subsection}{Subsection}{Subsections}

\title{Classification-Powered Conformal Inference for Zero-inflated Outcomes}

\author{
Zhirui Li$^{1}$,
Ricardo Diaz-Rincon$^{2}$,
Benjamin Shickel$^{3}$,
Sai Zhang$^{4}$,
Sohom Bhattacharya$^{5}$,
and Muxuan Liang$^{1}$\thanks{Corresponding author: \texttt{mliang2@mdanderson.org}} \\
$^{1}$Department of Biostatistics, The University of Texas MD Anderson Cancer Center, Houston, TX, USA \\
$^{2}$Department of Neuroscience, University of Florida, Gainesville, FL, USA \\
$^{3}$Department of Medicine, University of Florida, Gainesville, FL, USA \\
$^{4}$Department of Biomedical Informatics \& Data Science, Yale University, New Haven, CT, USA \\
$^{5}$Department of Statistics, University of Florida, Gainesville, FL, USA
}

\begin{document}
\maketitle

\begin{abstract}
Zero-inflated outcomes, where responses are zero with positive probability and otherwise continuous, are common in biomedical, environmental, and social science studies. We propose a conformal prediction based framework that provides distribution-free uncertainty quantification tailored to such outcomes. Standard conformal methods often ignore strong predictors distinguishing zero from non-zero outcomes, leading to overly conservative and unnecessarily long prediction sets. Our method integrates a classification step to identify zero outcomes and applies conformal inference to the non-zero part, producing prediction sets that are either ${0}$ or an interval. Under exchangeability, we establish that the proposed procedure attains the target marginal coverage and achieves asymptotically minimal interval length within this framework, regardless of the choice of classification or regression models. Extensive simulations and real-data application demonstrate the superior performance of our approach.
\end{abstract}


\maketitle

\section{Introduction}

Conformal inference is a general framework for constructing distribution-free prediction intervals for outcomes of interest based on predictors. A key advantage of conformal inference is that the resulting intervals achieve marginal coverage guarantees regardless of the underlying regression model. This property has made conformal methods widely used for uncertainty quantification in regression problems~\citep{vovk2005algorithmic, vovk2009line, JMLR:v9:shafer08a}. Previous works have explored its application with specific models, such as least squares and ridge regression \citep{burnaev2014efficiency}, nonparametric tolerance sets \citep{lei2013distribution}, nonparametric regression \citep{lei2014distribution}, and classification and clustering \citep{lei2014classification}. However, these works have primarily focused on continuous or categorical outcomes. 

In many practical settings, outcomes arise as mixtures of continuous and discrete components. A common example is zero-inflated data, where the outcome takes the value zero with a positive probability. Such data is prevalent across domains. In biology, species abundance data (e.g., gut microbiome profiles) often include absent species for many samples \citep{zeng2022mbdenoise}; in medical applications, treatment adjustments, e.g., dose changing, for chronic conditions may exhibit excess zeros, as dose changes are unnecessary unless symptoms are not controlled \citep{espay2017precision, bloem2021parkinson}. In these settings, zeros are not artifacts but carry critical practical meaning that directly informs decision-making.

Our motivating example is to predict the hourly averaged concentration of carbon monoxide ($\mathrm{CO}$), one of the most common air pollutants. Exposure to excessive $\mathrm{CO}$ is associated with respiratory symptoms, particularly among patients with chronic respiratory diseases \citep{song2023association, north2019personal}. Accurate forecasts of excessive concentrations beyond a pre-specified tolerance threshold can help guide patients in modifying their daily activities. Specifically, when the hourly averaged concentration remains below the tolerance, the outcome is recorded as $0$, indicating no warning. When it exceeds the tolerance, the outcome is defined as the concentration minus the tolerance, representing the magnitude of the exceedance and serving as an alert. Our goal is to construct informative prediction intervals, with guaranteed marginal coverage, for the excessive hourly averaged concentration of $\mathrm{CO}$ in order to support activity planning for at-risk populations.

For such zero-inflated outcomes, informative prediction requires two components: reliable prediction of zeros, and interpretable interval predictions for non-zero outcomes. In our example, accurate prediction of whether the concentration is within tolerance (zeros) guides patients’ activity decisions, while interval predictions in the exceedance regime quantify the severity of air pollution in an interpretable form. Importantly, interval-valued predictions are preferable to more complex sets, as they facilitate communication and decision-making \citep{lei2014classification,IntHout2016PI,GneitingRaftery2007,Winkler1972,sadinle2019least}. Although existing conformal inference methods can, in principle, be applied to zero-inflated outcomes \citep{vovk2005algorithmic, vovk2009line, JMLR:v9:shafer08a, burnaev2014efficiency, chernozhukov2021distributional}, they do not directly ensure informative predictions in this setting. In particular, they may fail to (i) provide singleton zero predictions when the outcome is likely to be zero, or (ii) return interval predictions when the outcome is non-zero. Addressing these limitations is the focus of our proposed framework (see Section~\ref{sec:related}).


In this work, we develop a two-step conformal inference procedure tailored to zero-inflated outcomes. In the first step, a classification model is used to distinguish between zero and non-zero outcomes. If the outcome is classified as zero, the prediction set is $\{0\}$. Otherwise, in the second step, a standard conformal method—such as split conformal inference \citep{vovk2005algorithmic}—is applied to construct an interval for the non-zero component. This yields prediction sets that are either a singleton $\{0\}$ or an interval, aligning with the natural structure of zero-inflated data. By decoupling the classification and regression tasks, our framework exploits strong predictors of zero outcomes to reduce the effective calibration burden. High accuracy in zero prediction relaxes the coverage demand on the non-zero regime (see Section~\ref{sec:rationale} for details), thereby enabling shorter and more efficient intervals while preserving marginal coverage guarantees. Furthermore, fitting the regression model only on the non-zero subset alleviates issues of zero inflation, improving both robustness and efficiency of the conformal procedure.

Our key contributions are summarized as the following. First, we introduce a new framework, \emph{Classification-Powered Conformal Inference (CPCI)}, designed for zero-inflated outcomes. This method integrates a classification model to predict zero outcomes and a regression model to construct conformal intervals for non-zero outcomes, yielding prediction sets that are either ${0}$ or a connected interval. Second, we establish that CPCI achieves the desired nominal marginal coverage under the sole assumption of exchangeability. The result holds for any classification or regression model, including nonparametric and black-box machine learning methods, thereby providing a general distribution-free guarantee. Further, we prove that, within our proposed framework, CPCI produces prediction sets with asymptotically minimal length. In particular, CPCI is guaranteed to perform at least as well as vanilla conformal inference and strictly improves efficiency when strong predictors of zeros exist. 

The rest of the paper is organized as follows. Section~\ref{sec:related} reviews related work in conformal inference, and explains why the existing methods does not suffice for our problem. Section~\ref{sec:method} introduces the proposed method. In Section~\ref{sec:theory}, we show that the proposed method achieves the nominal marginal coverage, and it leads to asymptotically minimal prediction interval length within our framework. In Sections~\ref{sec:sim} and~\ref{sec:real_data}, we conduct simulations and the application to our motivating example to show the performance of the proposed method. Finally, in Section~\ref{sec:diss}, we provide a discussion on future directions. 

\section{Preliminaries}\label{sec:related}

In this section, we review the key idea of conformal inference, summarize related work, and discuss the main challenge posed by zero-inflated outcomes.

\subsection{Recap: Conformal Inference}

We begin with vanilla conformal inference and explain why it is sub-optimal for zero-inflated outcomes, as it often produces prediction intervals that are unnecessarily wide. Standard conformal methods (e.g., split conformal inference \citep{lei2015conformal, lei2018distribution}) randomly partition the data into a training set and a calibration set~\citep{papadopoulos2002inductive}. On the training set, one fits a regression model $\widehat{f}$ to predict outcomes $Y$ from features $X$. The choice of $\widehat{f}$ is unrestricted, allowing the use of nonparametric or machine learning methods.

Given $\widehat{f}$, a conformity score $S(x,y)$ measures how well an observation $(x,y)$ conforms to the fitted model. A common choice is the absolute residual: $S(x, y) \;=\; \bigl|\,y - \widehat{f}(x)\bigr|$. Using the calibration set $\mathcal{D}_{\text{cal}} = \{(x_i, y_i)\}_{i=1}^m$, we compute the calibration scores $\{S_i\}_{i=1}^m$, where $S_i = S(x_i,y_i)$. Sorting these scores and taking the empirical quantile following $
  \widehat{Q}(\alpha) \;=\; \min\Bigl\{\, t : \sum_{i=1}^m \mathbb{I}\bigl\{S_i \le t\bigr\}\;\ge\; \lfloor(m+1)\alpha\rfloor \Bigr\}
$, we have the critical threshold.

Exchangeability of the calibration data and the new sample $(X_{\text{new}}, Y_{\text{new}})$ implies that
\begin{equation}\label{eq:coverage_key_idea}
\mathbb{P}\bigl(S(X_{\text{new}}, Y_{\text{new}})\leq \widehat{Q}(\alpha)\bigr) \geq \alpha.
\end{equation}
Inverting this condition yields the prediction interval $
  \mathcal{C}(X_{\text{new}})= 
  \Bigl[\widehat{f}(X_{\text{new}}) - \widehat{Q}(\alpha),
         \widehat{f}(X_{\text{new}})+ \widehat{Q}(\alpha)\Bigr].
$ Thus, vanilla conformal inference guarantees valid marginal coverage for any conformity score $S$ and any fitted model $\widehat{f}$. This distribution-free validity is its key strength. Several extensions have adapted the method to dependent or non-exchangeable data, including time series, online learning, and other structured settings \citep{barber2023conformal,wang2024online, gibbs2024conformal, oliveira2024split, zhou2024conformal, lee2025leave, leekernel}.

Although the coverage guarantee in~\eqref{eq:coverage_key_idea} holds for \emph{any} conformal score function $S$ and predictor $\widehat{f}$, the choice of $S$ and $\widehat{f}$ can have a substantial impact on the resulting interval length. In particular, commonly used scores—such as absolute residuals relative to conditional mean predictions—may fail to adequately capture structural features of the outcome distribution, for example the presence of excessive zeros. As a result, the constructed prediction intervals can become unnecessarily wide. Moreover, such choices may not yield singleton predictions of the form $\{0\}$, which are especially relevant in our applications.

\subsection{Related Work and Challenges}

\textbf{Modifying Conformal Score.} A natural extension of vanilla conformal prediction for zero-inflated outcomes is to define separate conformity scores for zero and non-zero outcomes:
$
S(x,y)=S^0 (x) 1\{y=0\}+S^1 (x,y) 1\{y\not=0\},
$
where $S^0 (x)$ and $S^1 (x,y)$ are conformity scores for the zero and non-zero parts, respectively. The prediction set is then constructed following the standard conformal procedure with inequality~\eqref{eq:coverage_key_idea}. Define $\widetilde{\mathcal{C}}(\alpha)=\left\{y: S^1 (X_{\text{new}},y)\leq \widehat{Q}(\alpha)\right\}$. The prediction set is 
$\left\{0 \right\}\cup \widetilde{\mathcal{C}}(\alpha)$
if $S^0 (X_{\text{new}})\leq \widehat{Q}(\alpha)$; and equals $\widetilde{\mathcal{C}}(\alpha)$ otherwise. This formulation parallels ideas from \emph{regression-as-classification} problem~\citep{guha2024conformal, sesia2021conformal, stewart2023regression}.

Despite its appeal, this approach has important drawbacks. First, the prediction set need not be an interval, which conflicts with the continuity of outcomes in our setting. For instance, if two distinct values of hourly $\textrm{CO}$ concentration are deemed possible, then all intermediate values should also be admissible. Second, the method may fail to output $\{0\}$ even when strong predictors of zero outcomes exist. A desirable approach should (i) allow zero to be predicted when justified by the data, and (ii) yield connected prediction sets, i.e., intervals.

Finally, the question of optimality arises: given a collection of models and conformity scores, one may seek the smallest valid conformal prediction set. A recent work~\citep{yang2025selection} address this by selecting among multiple ML models to minimize interval length. By contrast, our framework integrates two distinct models -- a classifier and a regressor -- applied sequentially, which ensures that prediction sets respect the required shape constraints.

\textbf{Distributional and Quantile Conformal Inference.}
Distributional conformal inference builds on models for conditional distributions, including quantile and distribution regressions, which can accommodate zero-inflated outcomes \citep{chernozhukov2021distributional}. However, when used to construct prediction intervals, these methods cannot produce $\{0\}$ as a prediction; instead, their intervals always contain $0$. Quantile-based conformal methods face the same limitation \citep{romano2019conformalized, alaa2023conformalized}.

\textbf{Class-Conditional Coverage Guarantees.}
Another line of work targets class-conditional coverage guarantees for zero and non-zero outcomes separately \citep{vovk2012conditional}. In the categorical setting, classwise conformal inference has been extensively studied \citep{lofstrom2015bias, shi2013applications, sun2017applying, hechtlinger2018cautious, guan2022prediction, sadinle2019least}. However, the resulting prediction sets need not be connected, and class-conditional coverage is often unnecessary when marginal coverage suffices.

\textbf{Challenge of Zero-Inflated Outcomes.}
Our goal is to design a conformal procedure tailored to zero-inflated outcomes, particularly when strong predictors of zero versus non-zero outcomes exist. To align prediction sets with the prevalence of zeros, the procedure must satisfy a shape constraint: (i) output $\{0\}$ when supported by the data, and (ii) otherwise output an interval, all while preserving coverage guarantees. Existing approaches do not meet this requirement. Our proposed Classification-Powered Conformal Inference (CPCI) addresses this gap by producing prediction sets of the form $\{0\}$ or an interval. Moreover, under suitable conditions, CPCI attains asymptotic marginal coverage and achieves shorter interval lengths asymptotically.

\section{Classification-Powered Conformal Inference}\label{sec:method}

\subsection{General Framework}\label{sec:rationale}

In this section, we introduce our proposed method and its implementation. CPCI is a two-step method that augments standard conformal inference with a preliminary classification step, thereby increasing efficiency while retaining valid coverage. For a new subject with covariates $X_{\text{new}}$, CPCI operates as follows. First, a classification model predicts whether the outcome is zero or non-zero. If the prediction is zero, the prediction set is $\{0\}$. Otherwise, a conformal interval is constructed using standard regression–based conformal inference. The general form of the resulting prediction set is
\begin{eqnarray}\label{eq:general_format}
    \mathcal{\widetilde{C}}(X_{\text{new}}) 
  \;=\;
  \begin{cases}
    \left\{0\right\}, 
    \text{ if ``predicted as zero"}; \\
    \mathcal{C}(X_{\text{new}}) , 
    \text{ otherwise},
  \end{cases}
\end{eqnarray}
where $\mathcal{C}(X_{\text{new}})$ denotes the conformal interval constructed in the second step with \textit{modified} coverage requirement. The proposed method is flexible and can incorporate any off-the-shelf black-box classification model.

The main methodological challenge lies in preserving marginal coverage despite the use of an arbitrary classification model. This requires a careful adjustment of coverage between the zero and non-zero components. 
For simplicity of illustration, we assume the outcomes are non-negatives. Let $Y \in \{0\} \cup \mathbb{R}_{>0}$ be an outcome with feature $X$. A zero-inflated outcome typically satisfies that $\mathbb{P}(Y=0\mid X)>0$.

The data is split into training, validation, and calibration sets. On the training dataset $\mathcal{D}_{\text{train}}$, we fit a classification model for
$
  {p}(x) \;=\; \mathbb{P}\{Y \not= 0 \mid X = x\},
$
which is used to predict whether the outcome of $X_{\text{new}}$ is zero or not. We denote the estimator as $\widehat{p}(x)$. On $\mathcal{D}_{\text{train}}$, we also fit a regression model $\widehat{f}$. This regression model can be fitted on only samples with $Y>0$; the marginal coverage guarantee holds regardless of how $f$ is trained. The calibration data is split into two-folds For the first fold $\mathcal{D}_{\text{cal,1}}$, we select a classification threshold $\alpha_{r}$  given a hyperparameter $r\in[0, 1]$, i.e.,
$
  \alpha_{r}
  \;=\;
  \mathrm{quantile}\Bigl(\{\widehat{p}(x_i) : (x_i,y_i)\in \mathcal{D}_{\text{cal,1}}\}\cup \{1\},\,r\Bigr).
$
A new subject is ``predicted as zero'' if $\widehat{p}(X_{\text{new}})\le \alpha_{r}$. Otherwise, conformal inference is applied for $X_{\text{new}}$ using the second fold $\mathcal{D}_{\text{cal,2}}$.

Suppose we aim at a prediction set with a marginal coverage of a pre-chosen level $\widetilde{\alpha}$. The key question is as follows: \textit{if we have that $\widehat{p}(X_{\text{new}})>\alpha_{r}$, what coverage should be imposed such that the overall coverage will achieve this nominal marginal coverage $\widetilde{\alpha}$?} To answer this question, we calculate the marginal coverage of our two-step prediction procedure via the following decomposition:
\begin{eqnarray}\nonumber
    \mathbb{P}(Y_{\text{new}}\in\widetilde{\mathcal{C}}(X_{\text{new}}))
    &=&\mathbb{P}(Y_{\text{new}}=0\mid \widehat{p}(X_{\text{new}})\le \alpha_{r})\mathbb{P}(\widehat{p}(X_{\text{new}})\le \alpha_{r})\\
    &&+\mathbb{P}(Y_{\text{new}}\in \mathcal{C}(X_{\text{new}})\mid \widehat{p}(X_{\text{new}})>\alpha_{r})\mathbb{P}(\widehat{p}(X_{\text{new}})> \alpha_{r}).\label{eq:coverage}
\end{eqnarray}
The first term corresponds to coverage among those predicted as zero; the second corresponds to coverage among predicted non-zeros. Equation~\eqref{eq:coverage} explains why classification can increase the power of conformal inference. First, when there exist strong predictors of zero vs non-zero outcomes, the coverage on ``predicted as zero'' is expected to be high; thus the coverage requirement on $\{\widehat{p}(X_{\text{new}})>\alpha_{r}\}$ can be relaxed, i.e., the prediction interval length on $\{\widehat{p}(X_{\text{new}})>\alpha_{r}\}$ can be shorter than the vanilla conformal inference approach. Second, it could be easier to construct a more efficient prediction interval since the proportion of zeros is expected to be lower on $\{\widehat{p}(X_{\text{new}})>\alpha_{r}\}$.

Specifically, we can estimate the requested coverage on $\{\widehat{p}(X_{\text{new}})>\alpha_{r}\}$. Due to the definition of $\alpha_{r}$ and the exchangeability between the testing sample and $\mathcal{D}_{\text{cal},1}$, we have that $\mathbb{P}(\widehat{p}(X_{\text{new}})\le \alpha_{r})\in [r, r+1/m]$, where $m$ denotes the sample sizes of the calibration datasets. Define $\beta=\mathbb{P}(Y_{\text{new}}=0\mid \widehat{p}(X_{\text{new}})\le \alpha_{r})$, which is known as the negative predictive value (NPV) in classification. On the validation dataset $\mathcal{D}_{\text{val}}$, we can estimate $\beta$ by the proportion of truly zero outcomes among ``predicted as zero'': 
\begin{equation}\label{eq:defn_hatbeta}
    \widehat{\beta}
  \;=\;
  \sum_{(x_i,y_i)\in \mathcal{D}_{\text{val}}} \mathbb{I}\{\widehat{p}(x_i)\le \alpha_{r} \;\text{and}\; y_i=0\}/(mr).
\end{equation}
Finally, to achieve a nominal marginal coverage of $\widetilde{\alpha}$, the conformal interval for non-zeros $\mathcal{C}(X_{\text{new}})$ must satisfy that
$
    \mathbb{P}(Y_{\text{new}}\in\mathcal{C}(X_{\text{new}})\mid \widehat{p}(X_{\text{new}})>\alpha_{r})\geq (\widetilde{\alpha} \;-\; r\widehat{\beta})/(\,1-r\,).
$

Then we can implement the vanilla conformal inference procedure using the second fold of the calibration data, $\mathcal{D}_{\text{cal,2}}$, with only samples satisfying $\widehat{p}(x) > \alpha_{r}$. On the second fold of the calibration set $\mathcal{D}_{\text{cal,2}}$ with only samples satisfying $\widehat{p}(x) > \alpha_{r}$, we define conformity scores
$
  S_i \;=\; \bigl|\,y_i - \widehat{f}(x_i)\bigr|.
$
We sort $\{S_i\}$ in ascending order and select
\[
  q_{r} 
  \;=\; 
  \mathrm{quantile}\!\Bigl(\{S_i\}\cup\{+\infty\},\;\,\gamma\Bigr), \quad
  \gamma 
  \;=\;
  \max\!\Bigl\{0,\;
      \min\!\Bigl\{1,\,
          (\widetilde{\alpha} \;-\; r\widehat{\beta})/(\,1-r\,)
      \Bigr\}
  \Bigr\}.
\]
The $\mathcal{C}(X_{\text{new}})$ is then $\bigl[\widehat{f}(X_{\text{new}}) - q_{r},\;\,\widehat{f}(X_{\text{new}}) + q_{r}\bigr]$. Overall, for a new test point $X_{\text{new}}$, our prediction interval is
\begin{align}\label{eq:cpci_final_form}
\widetilde{\mathcal{C}}(X_{\text{new}})
=
\begin{cases}
  \{0\}, & \widehat{p}(X_{\text{new}})\le \alpha_r;\\
  \bigl[\widehat{f}(X_{\text{new}})-q_r,\;\widehat{f}(X_{\text{new}})+q_r\bigr], & \text{otherwise}.
\end{cases}
\end{align}
With this implementation, vanilla conformal inference can be considered a special case of our framework with $r=0$ (i.e., $\gamma=0$). When $r=0$, all samples will be ``predicted as non-zeros'' and thus undergo a vanilla conformal inference procedure. In addition, although our second step adopts the vanilla conformal inference, any other conformal inference procedure that has a finite-sample marginal coverage guarantee can replace the vanilla conformal inference.

\begin{remark}
    The proposed method is designed to yield interval predictions or, more generally, connected prediction sets. In settings where the connectedness of prediction sets is not essential, alternative strategies such as direct modifications of the conformal score may be preferable. Nevertheless, our framework remains applicable in such cases and can be combined with these approaches to enhance their power while preserving valid coverage.
\end{remark}

\subsection{Choice of Hyperparameter for Minimal Interval Length}

In our general framework, the parameter $r$ is a pre-fixed hyperparameter. As we show in Section~\ref{sec:theory}, for any $r$, the proposed procedure leads to prediction intervals with guaranteed coverage. However, for different choices of $r$, the value of $\widehat{\beta}$ is different, and thus $q_r$ will be different, i.e., the length of the prediction interval depends on the choice of $r$. We can select the value of $r$ to achieve the \textit{minimal averaged interval length.}

The averaged interval length is defined as $2(1-r)q_r$, which is the product of the prediction interval length on $\left\{\widehat{p}(x) > \alpha_{r}\right\}$ and the proportion of $\left\{\widehat{p}(x) > \alpha_{r}\right\}$. When $\widehat{p}(x) \leq \alpha_{r}$, our prediction interval includes only a single point with a length of $0$. Other metrics can be chosen as the optimization target, e.g., the averaged interval length for non-zero predictions. In the online Supplementary Materials, we provide simulations using the averaged interval length for non-zero predictions as the optimization target to select $r$. In our implementation, to select $r$, we define a grid $\{r_k\}_{k=1}^K \subset [0,1]$ and, for each $r_k$, compute a threshold $\alpha_{r_k}$ from the validation dataset, $\mathcal{D}_{\text{val}}$, i.e.,
$
  \alpha_{r_k}
  \;=\;
  \mathrm{quantile}\Bigl(\{\widehat{p}(x_i) : (x_i,y_i)\in \mathcal{D}_{\text{cal,1}}\}\cup\{1\},\,r_k\Bigr).
$
We calculate the $\widehat{\beta}$ using the samples in $\mathcal{D}_{\text{val}}$, and thus the $q_{r_k}$.
Among all $r_k$'s, we select the one (denoted $\widehat{r}$) that minimizes the averaged interval length, i.e., $2(1-r)q_r$, or the averaged interval length for non-zero predictions, i.e., $q_r$. We denote the minimizer as $\widehat{r}$.

\section{Theoretical Guarantees}\label{sec:theory}
In this section, we establish \emph{finite-sample marginal coverage guarantees} and \emph{minimal interval length} properties for our proposed CPCI method. Specifically, we prove that the prediction intervals achieve the nominal coverage $\widetilde{\alpha}$ without imposing any distributional conditions. We focus on two types of coverage guarantees in finite sample:
\begin{eqnarray}\label{eq:coverge_marginal}
&&\mathbb{P}\bigl(Y_{\text{new}} \in \widetilde{C}(X_{\text{new}})\mid \mathcal{D}_{\text{train}}\bigr);\\\label{eq:coverge_marginal_val}
    &&\mathbb{P}\bigl(Y_{\text{new}} \in \widetilde{C}(X_{\text{new}})\mid \mathcal{D}_{\text{train}}, \mathcal{D}_{\text{valid}}\bigr).
\end{eqnarray}
Coverage guarantee~\eqref{eq:coverge_marginal} is defined over the validation and calibration datasets; coverage guarantee~\eqref{eq:coverge_marginal_val} further investigates the tail behavior of the impact of the validation datasets on the finite-sample coverage. Further, under mild smoothness assumptions on the residual distribution, we show that the resulting prediction intervals are asymptotically minimal. The proof of all theorems can be found in the online Supplementary Materials.


\begin{theorem}\label{thm:1}
Consider our proposed conformal prediction procedure~\eqref{eq:cpci_final_form} that, at a nominal coverage level \(\widetilde{\alpha}\) and any fixed $r$, produces predictive sets \(\widetilde{C}_n(X_{\text{new}})\) for a new testing sample $X_{\text{new}}$ independent of original data. 
Then, we have
$
\mathbb{P}\bigl(Y_{\text{new}} \in \widetilde{C}_n(X_{\text{new}})\mid \mathcal{D}_{\text{train}}\bigr) \geq \widetilde{\alpha}-1/m,
$
where recall that $m$ denotes the sample size of the calibration dataset.
\end{theorem}

Theorem~\ref{thm:1} provides the finite-sample coverage guarantees for the marginal coverage~\eqref{eq:coverge_marginal}. Since the marginal coverage is defined conditional on the training dataset, the result in Theorem~\ref{thm:1} applies to arbitrary classification model $\widehat{p}(x)$ and regression model $\widehat{f}(x)$ under any sample size of the training dataset $\mathcal{D}_{\text{train}}$. 
However, the result in Theorem~\ref{thm:1} does not imply coverage guarantee~\eqref{eq:coverge_marginal_val} due to the uncertainty of the validation dataset. To mitigate the impact of the uncertainty of the validation dataset, we introduce a modified estimator $\widehat{\beta}$ for $\beta$. Theorem~\ref{thm:2} shows that under a modified estimator based on $\widehat{\beta}$, we can achieve coverage guarantee~\eqref{eq:coverge_marginal_val} with a large probability. We emphasize that the result in Theorem~\ref{thm:2} holds in finite-sample.



\begin{theorem}\label{thm:2}
Replace the $\widehat{\beta}$ defined by~\eqref{eq:defn_hatbeta} with
$
    \widetilde{\beta}=\widehat{\beta}-Cr^{-1}({\log n}/{n})^{1/2},
$
where $C$ is any positive constant larger than $2$.
Then, under Assumptions in Theorem~\ref{thm:1}, when $n\geq 3$, with a probability at least $1-n^{-C^2/2+1}$, we have
$
\mathbb{P}\bigl(Y_{\text{new}} \in \widetilde{C}_n(X_{\text{new}})\mid \mathcal{D}_{\text{train}}, \mathcal{D}_{\text{val}}\bigr) \geq \widetilde{\alpha}-1/m.
$
\end{theorem}

Following Theorem~\ref{thm:2}, in our implementation, we use $\widetilde{\beta}$ to replace the $\widehat{\beta}$; and the results hold for any fixed $r$ and selected $\widehat{r}$. Compared with Theorem~\ref{thm:1}, $\widetilde{\beta}$ plays an essential role. The intuition is that the mean of $\widehat{\beta}$ is the true $\beta$; however, there is always a positive probability such that $\widehat{\beta}$ is larger than $\beta$, and thus leads to under-coverage. By adjusting $\widehat{\beta}$, i.e., using $\widetilde{\beta}$, we control the tail behavior induced by the uncertainty of the validation data and achieve 1) stronger coverage guarantees; 2) the ``post-selection'' coverage guarantees under the selected $\widehat{r}$. In our numerical studies, $C=5/2$ performed well.

Next, we show the \emph{minimal interval length} property for our proposed CPCI method. Notice that our prediction interval is either $\{0\}$ or $C(X_{\text{new}})$. Thus, the resulting interval length is either $0$ or $2q_r$. In this work, we consider the average interval length w.r.t. the calibration datasets $\mathcal{D}_{\text{cal},1}$, $\mathcal{D}_{\text{cal},2}$, and the testing samples. We denote the average interval length under $r$ as $L(r)$. By definition, we have that
$
    L(r)= \mathbb{E}\left[2 (1-r)q_r\mid \mathcal{D}_{\text{train}}\right],
$
where $2q_r$ is the interval length of $C(X_{\text{new}})$ and $(1-r)$ is the proportion of non-zero predictions, i.e., $\mathbb{P}(\widehat{p}(X_{\text{new}})> \alpha_{r})$. The minimal interval length is defined as the minimal interval length when we have $m$ samples in $\mathcal{D}_{\text{cal},1}$ and infinite samples in $\mathcal{D}_{\text{cal},2}$. Given $r>0$, the interval length when we have $m$ samples in $\mathcal{D}_{\text{cal},1}$ and infinite samples in $\mathcal{D}_{\text{cal},2}$ is 
$
    L^*(r)= \mathbb{E}\left[2(1-r)F^{-1}(\gamma) \mid \mathcal{D}_{\text{train}}\right],
$
where $F^{-1}$ is the quantile function of the residual for $Y\not=0$ and assumed to be differentiable. 

\begin{theorem}\label{thm:interval_length}
Under the assumptions in Theorem~\ref{thm:1}, we further assume that $L(r)$ is Lipschitz continuous, and $L^*(r)$ is differentiable with bounded derivatives. We also assume that $Y$ and $X$ are bounded. Then we have
$
  L(\widehat{r})
    \;-\;
    \inf_r L^*(r)
  \;\;\le\;\; O_p\left(n^{-1/2}+m^{-1/2}\right).
$
\end{theorem}
From Theorem~\ref{thm:interval_length}, as $\min\{n, m\}\to\infty$, our proposed method leads to prediction intervals that are asymptotically minimal. Since vanilla conformal inference procedure is a special case of the proposed method, our method can achieve at least comparable interval lengths to the vanilla conformal inference procedure.


\section{Simulations}\label{sec:sim}
In this section, we conduct simulations to compare our proposed CPCI and vanilla conformal inference procedures. We generate four predictors $(x_{1}, x_{2}, x_{3}, x_{4})$ i.i.d.\ from $\mathcal{N}(0,1)$. To generate an outcome with zero inflation, we adopt a two-step procedure. For each observation $i$, we first generate an underlying outcome $W_i$ following either a linear or a non-linear model with independent Gaussian noise; we then determine a quantile threshold $t = \mathrm{quantile}(\{W_i\},\,p)$ and threshold outcomes under $t$ to be $0$, where the parameter $p$ controls the fraction of zeros among all generated outcomes. In our main text, we fix $p = 0.75$ and $\widetilde{\alpha} = 0.90$. Detailed data generating mechanisms, experiments with different levels of $p$, coverage requirement $\widetilde{\alpha}$, more covariates, and optimizing objectives can be found in the online Supplementary Materials.

Under each data generation set-up, we vary the total sample size ($N$) among $1000, \, 2000, \, 3000,$ $4000,$ and $5000$. Then the resulting dataset is randomly and equally partitioned into a training dataset $\mathcal{D}_{\text{train}}$, a validation dataset $\mathcal{D}_{\text{val}}$, and two calibration sets $\mathcal{D}_{\text{cal,1}}$ and $\mathcal{D}_{\text{cal,2}}$. For methods that require only two split-datasets, we partition the entire data into a training and a single calibration set, and thus the efficiency loss of multiple splitting for our proposed methods is taken into account when comparing with other methods. An independent testing dataset $\mathcal{D}_{\text{test}}$ is also generated to calculate the coverage of the derived prediction intervals from different methods. We compare several methods for producing prediction intervals: (i) vanilla conformal inference with linear regression (VCI), (ii) VCI with XGBoost (VCI-XGB), (iii) VCI with support vector machines (VCI-SVM), (iv) CPCI with logistic regression and linear regression (CPCI), (v) CPCI with random forest for classification and XGBoost for regression (CPCI-XGB), (vi) CPCI with support vector machines for classification and regression (CPCI-SVM), (vii) Class-conditional conformal prediction (CLASS-COND) \citep{vovk2012conditional}, and (viii) VCI with weighted conformal score (WEIGHTED-VCI), where $S^0(x)$ is the estimated probability of non-zeros, and $S^{1}(x,y)$ is the absolute residual in the modified/weighted conformal score. For each method, we use the generated data and construct prediction intervals for the samples in the testing dataset. Then, we calculate and compare the coverages of the derived prediction intervals from different methods. We report average overall coverage, average interval length, proportion of predictions including $0$, and average interval length for predicted non-zeros. 
We repeat each setup 1000 times. 

\begin{figure}
    \centerline{%
    \includegraphics[width=0.8\linewidth]{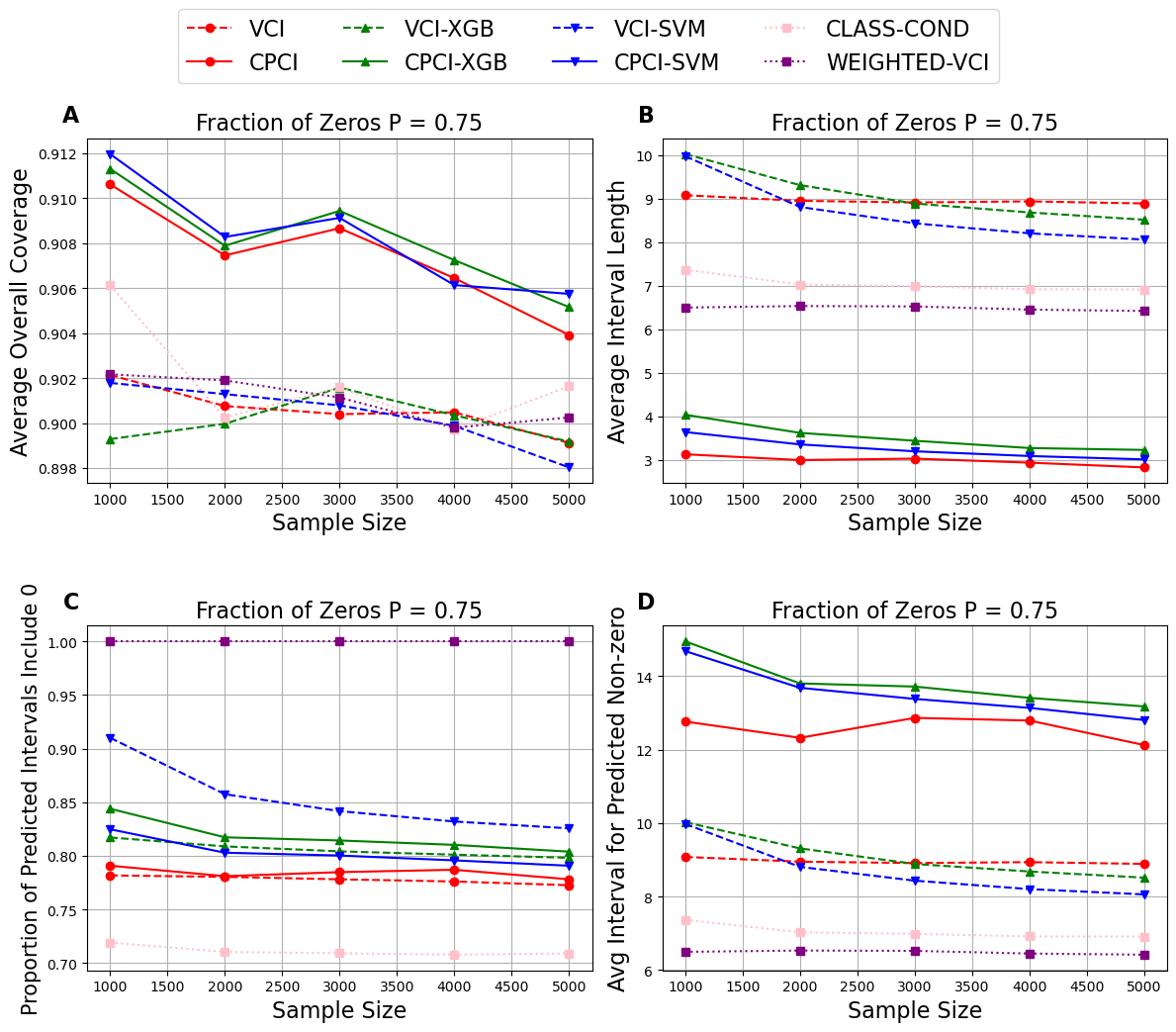}}
    \caption{\textbf{Linear Case Simulation Results.} 
   Subplots (A) show the marginal coverage, (B) the average interval length, (C) proportion of predicted intervals that include 0, and (D) the average interval length for predicted non-zero cases. At $\widetilde{\alpha}=0.90$, all three CPCI variants maintain nominal coverage and achieve significantly shorter average interval lengths than their VCI baselines, \textsc{CLASS-COND}, and \textsc{WEIGHTED-VCI}.}
    \label{fig_main:linear}
\end{figure}

\begin{figure}
    \centerline{%
    \includegraphics[width=0.8\linewidth]{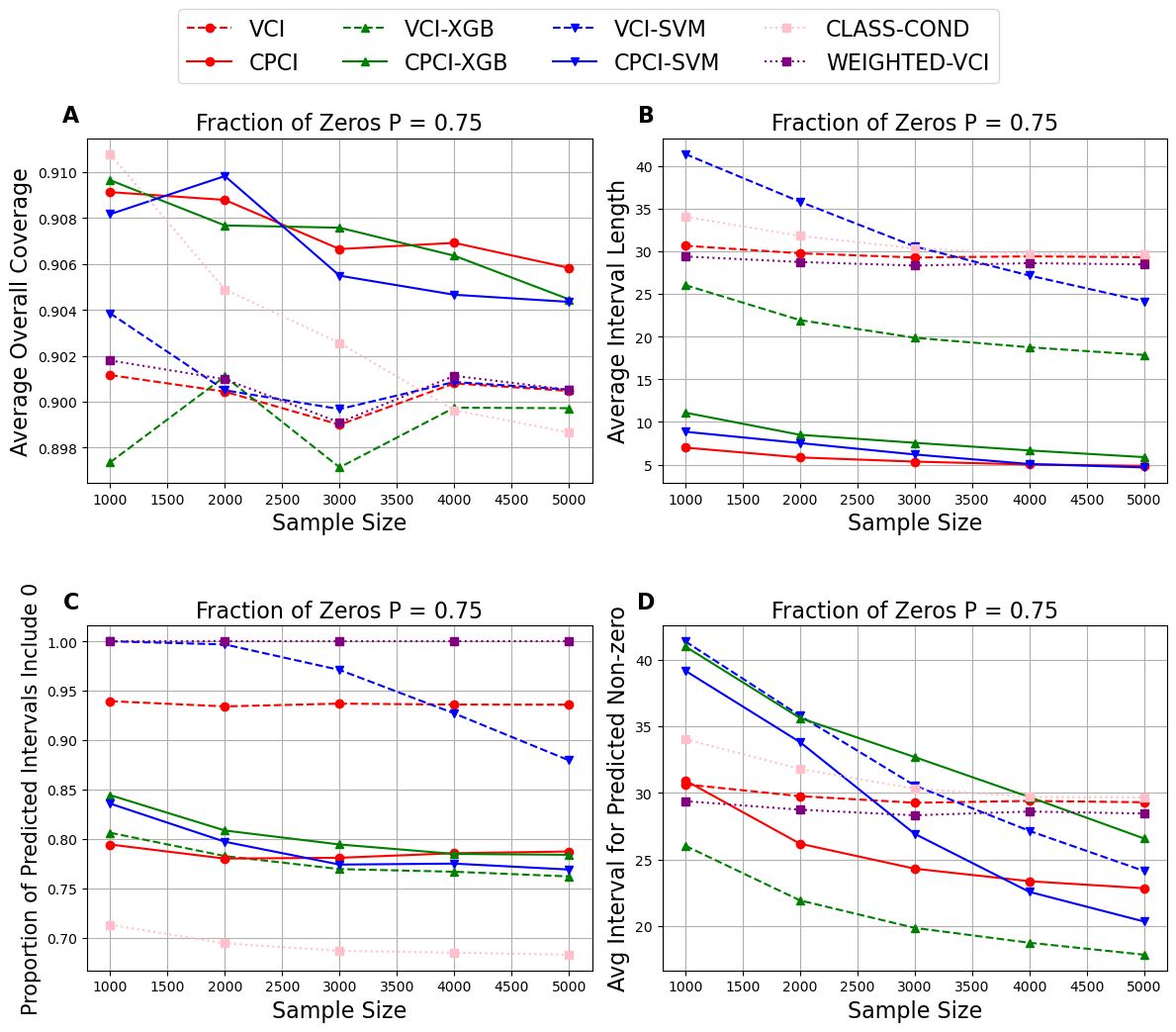}}
    \caption{ \textbf{Non-linear Case Simulation Results.} 
    Subplots (A) show the marginal coverage, (B) the average interval length, (C) proportion of predicted intervals that include 0, and (D) the average interval length for predicted non-zero cases.
    At $\widetilde{\alpha}=0.90$, all three CPCI variants maintain nominal coverage and achieve significantly shorter average interval lengths than their VCI baselines, \textsc{CLASS-COND}, and \textsc{WEIGHTED-VCI}. Moreover, \textsc{CPCI} and \textsc{CPCI-SVM} yield shorter average interval length for predicted non-zero cases.}
    \label{fig_main2:nonlinear}
\end{figure}

Figures~\ref{fig_main:linear} and~\ref{fig_main2:nonlinear} summarize the performance of the proposed and comparison methods. In the linear scenario, all approaches fluctuate around the specified nominal coverage $\widetilde{\alpha}=0.90$, but CPCI-based methods consistently yield significantly shorter average overall interval length. Overall, compared to VCI, the CPCI methods reduce the average interval length by more than $60\%$. 
Likewise, in the non-linear setting, all models center around 90\% coverage, with CPCI methods continuing to outperform vanilla conformal inference by producing significantly tighter intervals for average overall interval length. The CPCI methods achieve shorter prediction intervals than both CLASS-COND and WEIGHTED-VCI across all scenarios; in addition, both CLASS-COND and WEIGHTED-VCI produces disconnected prediction sets. In the online Supplementary Materials, we show that if we choose $r$ to minimize the interval length for non-zero predictions, our method also leads shorter or comparable average interval length for non-zero predictions when compared with other methods.

\section{Applications to Predicting Excessive Carbon Monoxide}\label{sec:real_data}

In this section, we apply our proposed method to the UCI Air Quality dataset. This dataset contains 9,358 hourly records collected by a five-sensor roadside air quality monitor in a heavily polluted Italian city between March 2004 and February 2005. Specifically, we focus on forecasting the hourly averaged concentration of $\textrm{CO}$ in ${mg}/{m^3}$. We consider multiple possible tolerance of negligible emissions or sensor thresholds—we set the response tolerance threshold as the 40th, 50th, 60th, 70th, or 80th quantile, i.e., the response is considered as $0$ if it falls below the tolerance threshold. In our implementation, we randomly partition the dataset into five equal subsets: $\mathcal{D}_{\text{train}}$, $\mathcal{D}_{\text{val}}$, $\mathcal{D}_{\text{cal,1}}$, $\mathcal{D}_{\text{cal,2}}$, and $\mathcal{D}_{\text{test}}$. For methods requiring only two splits, we merge $\mathcal{D}_{\text{val}}$, $\mathcal{D}_{\text{cal,1}}$, and $\mathcal{D}_{\text{cal,2}}$ into a single calibration set $\mathcal{D}_{\text{cal}}$, leaving $\mathcal{D}_{\text{train}}$ and $\mathcal{D}_{\text{test}}$ unchanged. We then implement the eight models detailed in Section~\ref{sec:sim} with a target coverage of $\widetilde{\alpha} = 0.9$. Additional evaluation metrics, results at other $\widetilde{\alpha}$ levels, and alternative optimization objectives are provided in the online Supplementary Materials.

Table~\ref{tab3:real} summarizes the performance of different methods. For zero proportions above \(50\%\), our proposed methods yield the minimum average interval length, with a reduction exceeding \(80\%\) relative to VCI. Although the CLASS-COND performs the best for Air Quality datasets with $40\%$ quantile as the threshold, it has 641.32 $\pm$ 20.85 (out of 1,535 test data) predictions as disconnected sets, i.e., the union of zero and an interval; and thus, the predictions are not informative in this case.

\begin{table}
\caption{UCI Air Quality dataset results at $\widetilde{\alpha}=0.90$.}
\label{tab3:real}
\begingroup
\setlength{\tabcolsep}{3pt} 
\footnotesize            
\begin{tabular*}{\columnwidth}{@{\extracolsep{\fill}}lcccccccc@{}}
\hline
 & VCI & VCI-XGB & VCI-SVM & CPCI & CPCI-XGB & CPCI-SVM & CLASS-COND & WEIGHTED-VCI \\
\hline
Tolerance & \multicolumn{8}{c}{\textbf{Average Overall Coverage}} \\
\hline
40\% & 0.90 $\pm$ 0.01 & 0.90 $\pm$ 0.01 & 0.90 $\pm$ 0.01 & 0.90 $\pm$ 0.01 & 0.90 $\pm$ 0.01 & 0.90 $\pm$ 0.01 & 0.90 $\pm$ 0.01 & 0.90 $\pm$ 0.01 \\
50\% & 0.90 $\pm$ 0.01 & 0.90 $\pm$ 0.01 & 0.90 $\pm$ 0.01 & 0.90 $\pm$ 0.01 & 0.90 $\pm$ 0.01 & 0.90 $\pm$ 0.01 & 0.90 $\pm$ 0.01 & 0.90 $\pm$ 0.01 \\
60\% & 0.90 $\pm$ 0.01 & 0.90 $\pm$ 0.01 & 0.90 $\pm$ 0.01 & 0.90 $\pm$ 0.01 & 0.90 $\pm$ 0.01 & 0.90 $\pm$ 0.01 & 0.90 $\pm$ 0.01 & 0.90 $\pm$ 0.01 \\
70\% & 0.90 $\pm$ 0.01 & 0.90 $\pm$ 0.01 & 0.90 $\pm$ 0.01 & 0.90 $\pm$ 0.01 & 0.90 $\pm$ 0.01 & 0.90 $\pm$ 0.01 & 0.90 $\pm$ 0.01 & 0.90 $\pm$ 0.01 \\
80\% & 0.90 $\pm$ 0.01 & 0.90 $\pm$ 0.01 & 0.90 $\pm$ 0.01 & 0.90 $\pm$ 0.01 & 0.90 $\pm$ 0.01 & 0.90 $\pm$ 0.01 & 0.90 $\pm$ 0.01 & 0.90 $\pm$ 0.01 \\
\hline
Tolerance & \multicolumn{8}{c}{\textbf{Average Interval Length}} \\
\hline
40\%  & 2.06 $\pm$ 0.04 & 1.94 $\pm$ 0.05 & 2.13 $\pm$ 0.06 & 1.84 $\pm$ 0.25 & 1.48 $\pm$ 0.26 & 2.00 $\pm$ 0.13 & \textbf{1.40 $\pm$ 0.04} & 1.61 $\pm$ 0.03 \\
50\%  & 2.34 $\pm$ 0.03 & 2.15 $\pm$ 0.06 & 2.39 $\pm$ 0.06 & 1.66 $\pm$ 0.27 & \textbf{1.09 $\pm$ 0.19} & 1.79 $\pm$ 0.26 & 1.44 $\pm$ 0.05 & 1.82 $\pm$ 0.04 \\
60\%  & 2.61 $\pm$ 0.03 & 2.32 $\pm$ 0.08 & 2.70 $\pm$ 0.07 & 1.09 $\pm$ 0.33 & \textbf{0.71 $\pm$ 0.12} & 1.11 $\pm$ 0.31 & 1.51 $\pm$ 0.07 & 2.12 $\pm$ 0.06 \\
70\%  & 2.81 $\pm$ 0.04 & 2.20 $\pm$ 0.11 & 2.87 $\pm$ 0.07 & 0.42 $\pm$ 0.06 & \textbf{0.39 $\pm$ 0.04} & 0.44 $\pm$ 0.07 & 1.60 $\pm$ 0.07 & 2.30 $\pm$ 0.06 \\
80\%  & 3.15 $\pm$ 0.04 & 2.25 $\pm$ 0.14 & 3.28 $\pm$ 0.09 & \textbf{0.19 $\pm$ 0.03} & 0.20 $\pm$ 0.02 & 0.21 $\pm$ 0.03 & 1.73 $\pm$ 0.10 & 2.34 $\pm$ 0.09 \\
\hline
\end{tabular*}
\endgroup
\bigskip
\end{table}

\section{Discussion}\label{sec:diss}

In this work, we propose a classification-powered conformal inference method (CPCI) for zero-inflated outcome prediction. The proposed method consists of two steps. The first step adopts a classification model to predict whether the outcome is zero or not; the second step adopts the vanilla conformal inference to construct a prediction interval. When there exist strong predictors of zero outcomes, CPCI can leverage the power of the strong predictors to shorten overall interval length or interval length for non-zero predictions. Under the exchangeability between the calibration datasets, validation datasets, and testing samples, we establish the the finite-sample marginal coverage guarantees for our proposed method under arbitrary classification and regression models. The resulting prediction interval length is (nearly) optimal compared to the case with infinite samples in the calibration datasets and, thus, at least comparable to the vanilla conformal inference. 


There are several possible future directions. First, in this work, we assume that the datasets are exchangeable with the testing samples. In practice, there could be distributional shifts between the testing samples, the validation data, and the calibration datasets. When there is a distributional shift, following \citep{NEURIPS2021_0d441de7,liu2024multi}, our framework can be extended. Second, although the proposed framework is designed for zero-inflated outcomes, the idea may apply to general outcomes. For example, suppose that the outcomes follow a mixture of multiple models; if there are strong predictors of which model the sample belongs to, our method may be able to increase the power of conformal inference by adaptively adjusting coverage requirements for each model such that the interval length can be minimized.

\section*{Supporting Information}

Web Appendices, referenced in Sections~\ref{sec:method} - ~\ref{sec:real_data}, and Python code implementing the proposed method are available with this paper at the Biometrics website on Oxford Academic. 
\vspace*{-8pt}

\section*{Data Availability}

The data that support this paper's findings are available publicly at UCI Machine Learning Repository.

\bibliographystyle{plainnat}
\bibliography{main}

\end{document}